\documentclass[preprint]{ptephy_v1}

\preprintnumber{KYUSHU-HET-189}

\usepackage{amsmath,amssymb}
\usepackage{mathtools}
\usepackage{booktabs}
\usepackage{subcaption}

\numberwithin{equation}{section}
\DeclareMathOperator{\sign}{sign}

\usepackage[%
pdftitle={Numerical study of the N=2 Landau--Ginzburg model with two superfields},%
pdfsubject={},%
pdfauthor={Okuto Morikawa},%
pdfkeywords={Last updated on \today%
}]{hyperref}

\title{Numerical study of the $\mathcal{N}=2$ Landau--Ginzburg model with two superfields}
\author{Okuto Morikawa}
\affil{Department of Physics, Kyushu University,
744 Motooka, Nishi-ku, Fukuoka, 819-0395, Japan
\email{o-morikawa@phys.kyushu-u.ac.jp}
}

\begin{document}
\begin{abstract}
In the low energy limit, the two-dimensional massless~$\mathcal{N}=2$
Wess--Zumino (WZ) model with a quasi-homogeneous superpotential
is believed to become a superconformal field theory.
This conjecture of the Landau--Ginzburg (LG) description has been
studied numerically in the case of the~$A_2$, $A_3$, and~$E_6$ minimal models.
In this paper, by using a supersymmetric-invariant non-perturbative formulation,
we simulate the WZ model with two superfields
corresponding to the~$D_3$, $D_4$, and~$E_7$ models.
Then, we numerically determine the central charge,
and obtain the results that are consistent with the conjectured correspondence.
We hope that this numerical approach, when further developed, will be useful
to investigate superstring theory via the LG/Calabi--Yau correspondence.
\end{abstract}
\subjectindex{B16, B24, B34}
\maketitle

\section{Introduction}
At an extremely low-energy scale, the two-dimensional massless~$\mathcal{N}=2$
Wess--Zumino (WZ) model~\cite{Wess:1974tw} with a quasi-homogeneous superpotential
is believed to become an~$\mathcal{N}=2$ superconformal field theory (SCFT).
This conjecture of the  Landau--Ginzburg (LG) description has been studied
from various aspects~\cite{DiVecchia:1985ief,DiVecchia:1986cdz,DiVecchia:1986fwg,%
Boucher:1986bh,Gepner:1986ip,Cappelli:1986hf,Cappelli:1986ed,Gepner:1986hr,%
Gepner:1987qi,Cappelli:1987xt,Kato:1987td,Gepner:1987vz,%
Kastor:1988ef,Vafa:1988uu,Martinec:1988zu,Lerche:1989uy,%
Howe:1989qr,Cecotti:1989jc,Howe:1989az,Cecotti:1989gv,Cecotti:1990kz,Witten:1993jg}.
However, we have no complete proof of this conjecture.
It is difficult to prove because
the coupling constant becomes strong at the low-energy region and
the perturbative theory possesses infrared (IR) divergences;
the LG description is a truly non-perturbative phenomenon.
An interesting approach to this issue may be a numerical and non-perturbative technique
on the basis of the lattice field theory.

By using such numerical approaches, the~$A_2$ and~$A_3$ minimal models were simulated
in~Refs.~\cite{Kawai:2010yj,Kamata:2011fr,Morikawa:2018ops},
where superpotentials in the corresponding WZ model
are given as the cubic and quartic ones containing a single superfield;
see~Table~\ref{tab:1}~\cite{Vafa:1988uu}.
These studies are based on either a lattice formulation in~Ref.~\cite{Kikukawa:2002as},
or a supersymmetry-preserving formulation with a momentum cutoff
in~Ref.~\cite{Kadoh:2009sp};
both non-perturbative formulations make essential use of the existence of the
Nicolai map~\cite{Nicolai:1979nr,Nicolai:1980jc,Parisi:1982ud,Cecotti:1983up}.\footnote{%
For some related works,
see~Refs.~\cite{Morikawa:2018ops,Kadoh:2009sp} and references therein.
It appears that the two-dimensional
massless~$\mathcal{N}=2$ WZ model is numerically studied in Ref.~\cite{Nicolis:2017lqk}.}
In above numerical studies,
their results of the scaling dimension and the central charge are consistent
with the expected values in the~$A_2$ and~$A_3$ minimal models within numerical errors.
Therefore, we have now numerical evidences for the LG/SCFT correspondence
in the case of the~$A_2$, $A_3$, and~$E_6$~($\cong A_2\otimes A_3$) minimal models.

\begin{table}[h]
 \centering
 \caption{ADE classification}
 \begin{tabular}{lll}\toprule
  Algebra & Superpotential $W$                         & Central charge $c$ \\\midrule
  $A_n$   & $\Phi^{n+1}$, $n\geqq1$                    & $3-6/(n+1)$        \\
  $D_n$   & $\Phi^{n-1}+\Phi\Phi^{\prime2}$, $n\geqq3$ & $3-6/2(n-1)$       \\
  $E_6$   & $\Phi^3+\Phi^{\prime4}$                    & $3-6/12$           \\
  $E_7$   & $\Phi^3+\Phi\Phi^{\prime3}$                & $3-6/18$           \\
  $E_8$   & $\Phi^3+\Phi^{\prime5}$                    & $3-6/30$           \\
  \bottomrule
 \end{tabular}
 \label{tab:1}
\end{table}

In this paper, on the basis of the momentum cutoff regularization~\cite{Kadoh:2009sp}
and the analysis in~Ref.~\cite{Morikawa:2018ops},
we simulate the two-dimensional $\mathcal{N}=2$ WZ model
corresponding to $D$- and $E$-type theories.
The method in~Ref.~\cite{Morikawa:2018ops} is generalized
to the WZ model with multiple superfields and more complicated superpotentials.
Then, from an IR behavior of the energy-momentum tensor (EMT),
we numerically determine the central charge of the~$D_3$, $D_4$, and~$E_7$ models;
we obtain the results that are consistent with the conjectured correspondence.
We also measure the ``effective central charge''~\cite{Kamata:2011fr,Morikawa:2018ops},
which is analogous to the
Zamolodchikov's~$c$-function~\cite{Zamolodchikov:1986gt,Cappelli:1989yu}.
Although the theoretical background of the formulation is not completely obvious so far,
our computational results support the validity
of the formulation even if we consider multi-superfield theories.
We hope to apply this approach to some models which is neither a minimal model
nor a product of minimal models (Gepner model~\cite{Gepner:1987qi,Gepner:1987vz}),
and then develop a numerical method to investigate superstring theory
via the LG/Calabi--Yau correspondence~\cite{Martinec:1988zu,%
Cecotti:1990wz,Greene:1988ut,Witten:1993yc}.

\section{Nicolai mapping for the multi-superfield WZ model}
Our numerical simulation is based on the formulation in~Ref.~\cite{Kadoh:2009sp}.
The detailed discussions for the formulation are given in~Ref.~\cite{Morikawa:2018ops}.
These preceding studies treat the two-dimensional~$\mathcal{N}=2$ WZ model
with a single superfield.
In this section, we summarize basic formulas of the formulation
for the WZ model with multiple superfields.

Suppose that the system is defined
in a two-dimensional Euclidean physical box~$L_0\times L_1$.
In what follows, we work in the momentum space with an ultraviolet (UV) cutoff,
\begin{align}
 p_\mu = \frac{2\pi}{L_\mu} n_\mu \qquad
 \left(n_\mu=0,\pm1,\pm2,\dots,\pm\frac{L_\mu}{2a}\right),
\end{align}
where the Greek index~$\mu$ runs over~$0$ and~$1$,
and repeated indices are not summed over;
$L_\mu/a$ is taken as even integers, and $a$ is a unit of dimensionful quantities.
A limit~$a\to0$ removes the UV cutoff, being similar to the \textit{continuum limit}.
In fact, when we take~$L_\mu/a$ as \textit{odd} integers,
the unit $a$ itself is the lattice spacing in the dimensional reduction
of the four-dimensional lattice formulation~\cite{Bartels:1983wm}
based on the SLAC derivative~\cite{Drell:1976bq,Drell:1976mj}.

It is well recognized that
the regularization based on the SLAC derivative violates the locality.
The four-dimensional SLAC derivative is plagued by the pathology
that the locality of the theory is not automatically restored
in the continuum limit~\cite{Dondi:1976tx,Karsten:1979wh,Kato:2008sp,Bergner:2009vg}.
In the two- or three-dimensional case, on the other hand,
one can argue the restoration of the locality in the continuum limit
within perturbation theory for \textit{massive} WZ models~\cite{Kadoh:2009sp}.
For \textit{massless} models,
it is not clear whether the restoration is automatically accomplished
because perturbation techniques are hindered by the IR divergences.
We believe that the numerical results in~Refs.~\cite{Kamata:2011fr,Morikawa:2018ops}
and ours below support the validity of the formulation.

For simplicity, we set~$a=1$. We basically use the complex coordinates for the momentum,
$p_z = (p_0-ip_1)/2$ and $p_{\Bar{z}} = (p_0+ip_1)/2$.
In general, the two-dimensional~$\mathcal{N}=2$ WZ model contains
$N_{\Phi}$ superfields, $\{\Phi_I\}_{I=1,\dots,N_{\Phi}}$.
A supermultiplet~$\Phi_I$ consists of a complex scalar~$A_I$,
left- and right-handed spinors~$(\psi_I, \Bar\psi_I)$,
and a complex auxiliary field~$F_I$.
Then, the action of the two-dimensional~$\mathcal{N}=2$ WZ model
with a quasi-homogeneous superpotential~$W(\{A\})$ is given by
\begin{align}
 S &= \frac{1}{L_0L_1} \sum_p \sum_I\Biggl[
 4 p_z A_I^*(-p) p_{\Bar{z}} A_I(p)
 - F_I^*(-p) F_I(p)
 \notag\\&\qquad\qquad\qquad\qquad
 - F_I(-p) \frac{\partial W(\{A\})}{\partial A_I}(p)
 - F_I^*(-p) \frac{\partial W(\{A\})^*}{\partial A_I^*}(p)
 \notag\\&\qquad\qquad\qquad\qquad
 + (\Bar\psi_{\Dot{1}}, \psi_2)_I(-p) \sum_J
 \begin{pmatrix}
  2\delta_{IJ}p_z & \frac{\partial^2 W(\{A\})^*}{\partial A_I^* \partial A_J^*}* \\
  \frac{\partial^2 W(\{A\})}{\partial A_I \partial A_J}* & 2\delta_{IJ}p_{\Bar{z}}
 \end{pmatrix}
 \begin{pmatrix}
  \psi_1 \\ \Bar\psi_{\Dot{2}}
 \end{pmatrix}_J(p)
 \Biggr] ,
 \label{eq:form.2}
\end{align}
where $*$ denotes the convolution
\begin{align}
 (\varphi_1 * \varphi_2)(p)
 \equiv \frac{1}{L_0L_1} \sum_q \varphi_1(q) \varphi_2(p-q).
\end{align}
The field products in~$\partial W(\{A\})/\partial A_I$
and~$\partial W(\{A\})/\partial A_I \partial A_J$ are understood as this convolution.
Integrating over the auxiliary fields~$\{F\}$,
we obtain the action in terms of the physical component fields,
\begin{align}
 S &= S_B + \frac{1}{L_0L_1} \sum_p \sum_{I,J}
 (\Bar\psi_{\Dot{1}}, \psi_2)_I(-p)
 \begin{pmatrix}
  2i\delta_{IJ}p_z & \frac{\partial^2 W(\{A\})^*}{\partial A_I^* \partial A_J^*}* \\
  \frac{\partial^2 W(\{A\})}{\partial A_I \partial A_J}* & 2i\delta_{IJ} p_{\Bar{z}}
 \end{pmatrix}
 \begin{pmatrix}
  \psi_1 \\ \Bar\psi_{\Dot{2}}
 \end{pmatrix}_J(p) ,
\end{align}
where $S_B$ is the bosonic part of the total action
\begin{align}
 S_B &= \frac{1}{L_0L_1} \sum_p \sum_I N_I^*(-p) N_I(p) , &
 N_I(p) &= 2p_z A_I(p) + \frac{\partial W(\{A\})^*}{\partial A_I^*}(p) .
 \label{eq:form.5}
\end{align}

The new variables $\{N\}$~\eqref{eq:form.5} specify the so-called Nicolai
map~\cite{Nicolai:1979nr,Nicolai:1980jc,Parisi:1982ud,Cecotti:1983up},
the change of variables from~$\{A\}$ to~$\{N\}$.
This mapping simplifies the path-integral weight drastically;
the partition function where the fermion fields are integrated is given by
\begin{align}
 \mathcal{Z}
 &= \int \prod_{|p_\mu|\leq\pi} \prod_I [d N_I(p) d N_I^*(p)]\,
 e^{- S_B}
 \sum_k \left.\sign \det
 \frac{\partial(\{N\}, \{N^*\})}{\partial(\{A\}, \{A^*\})} \right|_{\{A\}=\{A\}_{k}},
\end{align}
where~$\{A\}_{k}$ ($k=1$, $2$, \dots) is a set of solutions of the equation
\begin{align}
 2i p_z A_I(p) + \frac{\partial W(\{A\})^*}{\partial A_I^*}(p) - N_I(p) = 0.
 \label{eq:form.7}
\end{align}
The weight~$\exp(-S_B)$ is a Gaussian function of the variables~$\{N\}$.
All we should do to obtain configurations of~$\{N\}$ and~$\{A\}$ is the following:
We generate complex random numbers~$\{N(p)\}$ from the Gaussian distribution
for each momentum, and then,
solve numerically the algebraic equation~\eqref{eq:form.7} with respect to~$\{A\}$.
Note that, however, our numerical root-finding analysis may suffer from the
systematic error associated with some undiscovered solutions to~Eq.~\eqref{eq:form.7}.
Ref.~\cite{Morikawa:2018ops} addresses the difficulty of the algorithm.

\section{Central charge from the EMT correlator}
In a two-dimensional SCFT,
the central charge~$c$ appears in the two-point function of EMT;
in terms of the Fourier mode~$T(p)\equiv T_{zz}(p)$,
we have~\cite{Morikawa:2018ops}\footnote{In the spacetime
with the complex coordinates~$z\equiv x_0+ix_1$ and~$\Bar{z}\equiv x_0-ix_1$,
the EMT correlator is given by~$\langle T(z) T(0) \rangle = c/2z^4$.
This convention is identical
to that of~Refs.~\cite{Morikawa:2018ops,Polchinski:1998rq,Polchinski:1998rr}.
}
\begin{align}
 \langle T(p) T(-p) \rangle
 &= L_0L_1 \frac{\pi c}{12} \frac{p_z^3}{p_{\Bar{z}}}.
 \label{eq:emt.2}
\end{align}
The two-dimensional~$\mathcal{N}=2$ WZ model, which itself is not
superconformal invariant and hence does not behave as~Eq.~\eqref{eq:emt.2},
is believed to give an LG description of SCFT.
Thus, the IR behavior of the WZ model would be governed
by relations as~Eq.~\eqref{eq:emt.2} in SCFT.
The central charge can be computed
from the fit function~\eqref{eq:emt.2} in the IR region.

Let us write down explicit expressions of EMT and its correlator
in the WZ model~\eqref{eq:form.2}.
Since our formulation preserves some spacetime symmetries exactly,
we can straightforwardly construct Noether currents associated with the symmetries,
for example, the supercurrent and EMT.
To remove the ambiguity of EMT, we require the traceless
condition~$T_{z\Bar{z}}=T_{\Bar{z}z}=0$ in the free-field limit~\cite{Morikawa:2018ops}
(see also~Refs.~\cite{Tseytlin:1987bz,Polchinski:1987dy}).
Following the corresponding computation in~Ref.~\cite{Morikawa:2018ops},
EMT is given in the momentum space by
\begin{align}
 T(p)
 &= \frac{\pi}{L_0L_1} \sum_q \sum_I \Bigl[
 4 (p-q)_z q_z A_I^*(p-q) A_I(q) \notag\\
 &\qquad\qquad\qquad\quad
 - i q_z \psi_{2I}(p-q) \Bar\psi_{\Dot{2}I}(q)
 + i (p-q)_z \psi_{2I}(p-q) \Bar\psi_{\Dot{2}I}(q)
 \Bigr] .
\end{align}
Like as the single-supermultiplet case, it can turn out that this expression of EMT is
the super-transformation of the supercurrent;
for the definition of the super-transformation,
see Appendix~A in~Ref.~\cite{Morikawa:2018ops}.
Thus we can obtain a less noisy form of the EMT correlator
\begin{align}
 \left\langle T(p) T(-p) \right\rangle
 &= - \frac{1}{16} 2i p_z 
 \left\langle
 S_z^{-}(p) S_z^{+}(-p)
 + S_z^{+}(p) S_z^{-}(-p)
 \right\rangle ,
 \label{eq:emt.4}
\end{align}
where $S_z^{\pm}$ is the supercurrent defined by
\begin{align}
 S_z^{+}(p)
 &= \frac{4\pi}{L_0L_1}  \sum_q \sum_I i(p-q)_z A_I(p-q) \Bar{\psi}_{\Dot{2}I}(q) ,\\
 S_z^{-}(p)
 &= - \frac{4\pi}{L_0L_1}\sum_q \sum_I i(p-q)_z A_I^*(p-q) \psi_{2I}(q) .
\end{align}
Since the formulation exactly preserves the supersymmetry,
this relation between EMT and the supercurrent holds.

\section{Numerical setup and sampling configurations}
In what follows, we consider the~$N_{\Phi}=2$ WZ model
corresponding to the~$D_3$, $D_4$, and~$E_7$ minimal models.
The superpotential is defined by
\begin{align}
 W(\{\Phi\}) &= \frac{\lambda_1}{n} \Phi_1^n + \frac{\lambda_2}{2} \Phi_1 \Phi_2^2
 \qquad \text{for}\, D_n, \\
 W(\{\Phi\}) &= \frac{\lambda_1}{3} \Phi_1^3 + \frac{\lambda_2}{3} \Phi_1 \Phi_2^3
 \qquad \text{for}\, E_7.
\end{align}
We set the couplings~$\lambda_1$ and~$\lambda_2$ to~$0.3$ in the unit of~$a=1$.
The box size~$L = L_0 = L_1$ is taken as
even integers~$8$, $16$, $24$, $32$, $40$, and~$44$ for~$D_3$;
$8$, $16$, $24$, $32$, $36$, $40$ and~$42$ for~$D_4$; $8$, $16$, and~$24$ for~$E_7$.
Other numerical setups are identical to those in~Ref.~\cite{Morikawa:2018ops},
and hence see~Ref.~\cite{Morikawa:2018ops} for more detailed information of our program.

The classification of obtained configurations are tabulated
in~Tables~\ref{tab:class.spt-d_lm0-0.300000lm1-0.300000k2L8_1}--%
\ref{tab:class.spt-e_lm0-0.300000lm1-0.300000k3L8_1}.
We also listed the Witten index~\cite{Witten:1982df,Cecotti:1981fu,Morikawa:2018ops},
\begin{align}
 \Delta \equiv
 \left\langle \left.
 \sum_k\sign\det\frac{\partial(\{N\},\{N^*\})}{\partial(\{A\},\{A^*\})}
 \right|_{\{A\}=\{A\}_{k}} \right\rangle
 = n \quad\text{for}\, D_n, E_n,
 \label{eq:set.3}
\end{align}
and the one-point
supersymmetry Ward--Takahashi identity~\cite{Catterall:2001fr,Morikawa:2018ops}
\begin{align}
 \delta \equiv \frac{\left\langle S_B \right\rangle}{(L+1)^2} - 1 = 0.
 \label{eq:set.4}
\end{align}
$\Delta$ and~$\delta$ should be reproduced numerically,
and indicate the quality of our configurations.
The computational time is given in core$\cdot$hour
on the Intel Xeon E5 2.0 GHz for~$D_3$ with~$L=8$, \dots, $44$
and~$D_4$ with~$L=8$, \dots, $32$, $40$
and the Intel Xeon Gold 3.0 GHz for~$D_4$ with~$L=36$, $42$
and~$E_7$ with~$L=8$, \dots, $24$.

\begin{table}[ht]
 \centering
 \caption{Classification of configurations for~$D_{3}$.
The symbol~$({+}\dots{+}{-}\dots{-})_{n_{+}-n_{-}}$
with $n_{+}$ plus and~$n_{-}$ minus symbols implies that, for a configuration~$\{N(p)\}$,
we find $(n_{+}+n_{-})$~solutions, $\{A(p)\}_k$ ($k=1$, \dots, $n_{+}+n_{-}$);
the~$n_{+}$ solutions take $\sign\det\frac{\partial(N,N^*)_i}{\partial(A,A^*)_j}={+}1$
and the~$n_{-}$ solutions take~${-}1$.
The number of obtained configurations for each setup is shown.
$\Delta$~\eqref{eq:set.3} and~$\delta$~\eqref{eq:set.4} are numerically computed.
 }
 \begin{tabular}{lrrr}\toprule
  $L$         & 8 & 16 & 24 \\\midrule
  $({+}{+}{+})_3$ & 640 & 640 & 638 \\
  $({+}{+}{+}{+}{-})_3$ & 0 & 0 & 2 \\
  $({+}{+}{+}{+}{+}{-}{-})_3$ & 0 & 0 & 0 \\
  \midrule
  $\Delta$ & 3 & 3 & 3 \\
  $\delta$ & 0.0016(30) & 0.0015(16) & $-$0.0007(11) \\
  \midrule
  core$\cdot$hour [h] & 1.60 & 63.78 & 649.73 \\
  \bottomrule
 \end{tabular}
 \label{tab:class.spt-d_lm0-0.300000lm1-0.300000k2L8_1}
\end{table}%
\begin{table}[ht]
 \centering
 \caption{Classification of configurations for~$D_{3}$ (continued).}
 \begin{tabular}{lrrr}\toprule
  $L$         & 32 & 40 & 44 \\\midrule
  $({+}{+}{+})_3$ & 639 & 633 & 632 \\
  $({+}{+}{+}{+}{-})_3$ & 1 & 7 & 7 \\
  $({+}{+}{+}{+}{+}{-}{-})_3$ & 0 & 0 & 1 \\
  \midrule
  $\Delta$ & 3 & 3 & 3 \\
  $\delta$ & 0.00019(85) & $-$0.00078(69) & $-$0.00092(63) \\
  \midrule
  core$\cdot$hour [h] & 3440.33 & 13426.08 & 25623.00 \\
  \bottomrule
 \end{tabular}
 \label{tab:class.spt-d_lm0-0.300000lm1-0.300000k2L8_2}
\end{table}%
\begin{table}[ht]
 \centering
 \caption{Classification of configurations for~$D_{4}$.
 Same as Table~\ref{tab:class.spt-d_lm0-0.300000lm1-0.300000k2L8_1}, but for~$D_{4}$.}
 \begin{tabular}{lrrrr}\toprule
  $L$         & 8 & 16 & 24 & 32 \\\midrule
  $({+}{+}{+}{+})_4$ & 640 & 638 & 629 & 626 \\
  $({+}{+}{+}{+}{+}{-})_4$ & 0 & 2 & 9 & 13 \\
  $({+}{+}{+}{+}{+}{+}{-}{-})_4$ & 0 & 0 & 2 & 1 \\
  $({+}{+}{+}{+}{+}{+}{+}{-}{-}{-})_4$ & 0 & 0 & 0 & 0 \\
  $({+}{+}{+}{+}{-})_3$ & 0 & 0 & 0 & 0 \\
  $({+}{+}{+}{+}{+})_5$ & 0 & 0 & 0 & 0 \\
  $({+}{+}{+}{+}{+}{+}{+}{+}{+}{+}{-}{-}{-}{-}{-}{-})_4$ & 0 & 0 & 0 & 0 \\
  \midrule
  $\Delta$ & 4 & 4 & 4 & 4 \\
  $\delta$ & $-$0.0028(31) & 0.0018(17) & $-$0.0020(11) & $-$0.00095(84) \\
  \midrule
  core$\cdot$hour [h] & 1.73 & 72.28 & 847.83 & 5220.60 \\
  \bottomrule
 \end{tabular}
 \label{tab:class.spt-d_lm0-0.300000lm1-0.300000k3L8_1}
\end{table}%
\begin{table}[ht]
 \centering
 \caption{Classification of configurations for~$D_{4}$ (continued).}
 \begin{tabular}{lrrr}\toprule
  $L$         & 36 & 40 & 42 \\\midrule
  $({+}{+}{+}{+})_4$ & 604 & 606 & 603 \\
  $({+}{+}{+}{+}{+}{-})_4$ & 23 & 24 & 35 \\
  $({+}{+}{+}{+}{+}{+}{-}{-})_4$ & 12 & 8 & 9 \\
  $({+}{+}{+}{+}{+}{+}{+}{-}{-}{-})_4$ & 1 & 0 & 0 \\
  $({+}{+}{+}{+}{-})_3$ & 0 & 1 & 0 \\
  $({+}{+}{+}{+}{+})_5$ & 0 & 1 & 0 \\
  $({+}{+}{+}{+}{+}{+}{+}{+}{+}{+}{-}{-}{-}{-}{-}{-})_4$ & 0 & 0 & 1 \\
  \midrule
  $\Delta$ & 4 & 4.000(2) & 4 \\
  $\delta$ & $-$0.00008(73) & $-$0.00025(87) & 0.00019(64) \\
  \midrule
  core$\cdot$hour [h] & 4328.58 & 22264.12 & 12272.93 \\
  \bottomrule
 \end{tabular}
 \label{tab:class.spt-d_lm0-0.300000lm1-0.300000k3L8_2}
\end{table}%
\begin{table}[ht]
 \centering
 \caption{Classification of configurations for~$E_{7}$.
 Same as Table~\ref{tab:class.spt-d_lm0-0.300000lm1-0.300000k2L8_1}, but for~$E_{7}$.}
 \begin{tabular}{lrrr}\toprule
  $L$         & 8 & 16 & 24 \\\midrule
  $({+}{+}{+}{+}{+}{+}{+})_7$ & 639 & 628 & 582 \\
  $({+}{+}{+}{+}{+}{+}{+}{+}{-})_7$ & 1 & 10 & 20 \\
  $({+}{+}{+}{+}{+}{+})_6$ & 0 & 2 & 27 \\
  $({+}{+}{+}{+}{+}{+}{+}{-})_6$ & 0 & 0 & 5 \\
  $({+}{+}{+}{+}{+}{+}{+}{+})_8$ & 0 & 0 & 3 \\
  $({+}{+}{+}{+}{+}{+}{+}{+}{-}{-})_6$ & 0 & 0 & 2 \\
  $({+}{+}{+}{+}{+}{+}{+}{+}{+}{-}{-})_7$ & 0 & 0 & 1 \\
  \midrule
  $\Delta$ & 7 & 6.997(2) & 6.952(9) \\
  $\delta$ & $-$0.0013(32) & $-$0.0004(16) & 0.0009(17) \\
  \midrule
  core$\cdot$hour [h] & 1.30 & 60.28 & 750.92 \\
  \bottomrule
 \end{tabular}
 \label{tab:class.spt-e_lm0-0.300000lm1-0.300000k3L8_1}
\end{table}%

\clearpage
\section{Numerical determination of the central charge}
Let us show the main result of this paper,
the numerical determination of the central charge for the~$D_3$, $D_4$, and~$E_7$ models.

\begin{figure}[ht]
 \begin{center}
  \begin{subfigure}{0.48\columnwidth}
   \begin{center}
    \includegraphics[width=\columnwidth]{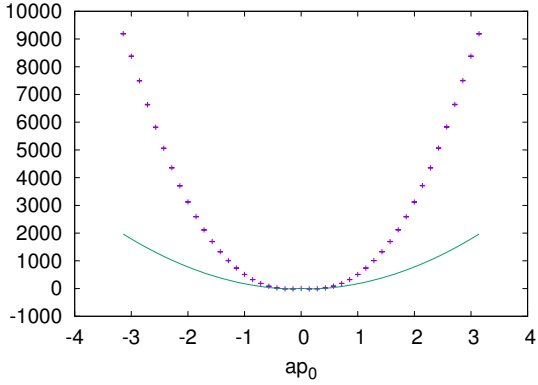}
    \caption{Real part}
   \end{center}
  \end{subfigure} \hspace*{1em}
  \begin{subfigure}{0.48\columnwidth}
   \begin{center}
    \includegraphics[width=\columnwidth]{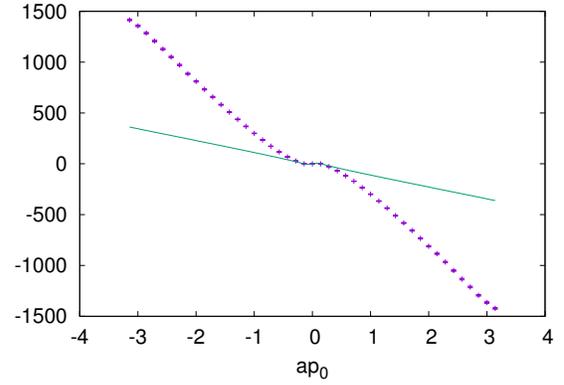}
    \caption{Imaginary part}
   \end{center}
  \end{subfigure}
 \end{center}
 \caption{$\langle T(p)T(-p)\rangle$ for~$D_3$, $L=44$, and~$p_1=\pi/22$.
 The fitting curve~\eqref{eq:emt.2} is depicted at once.}
 \label{fig:emt_spt-d_k2L44}
\end{figure}

\begin{figure}[ht]
 \begin{center}
  \begin{subfigure}{0.48\columnwidth}
   \begin{center}
    \includegraphics[width=\columnwidth]{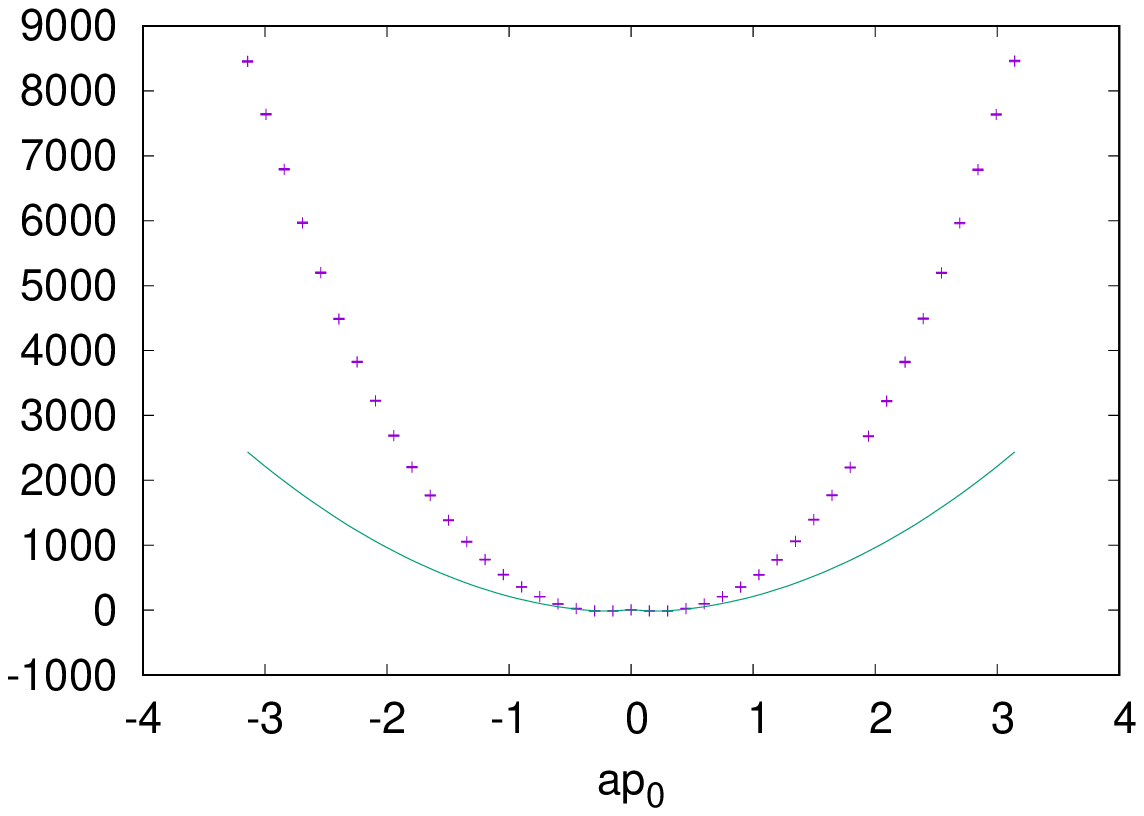}
    \caption{Real part}
   \end{center}
  \end{subfigure} \hspace*{1em}
  \begin{subfigure}{0.48\columnwidth}
   \begin{center}
    \includegraphics[width=\columnwidth]{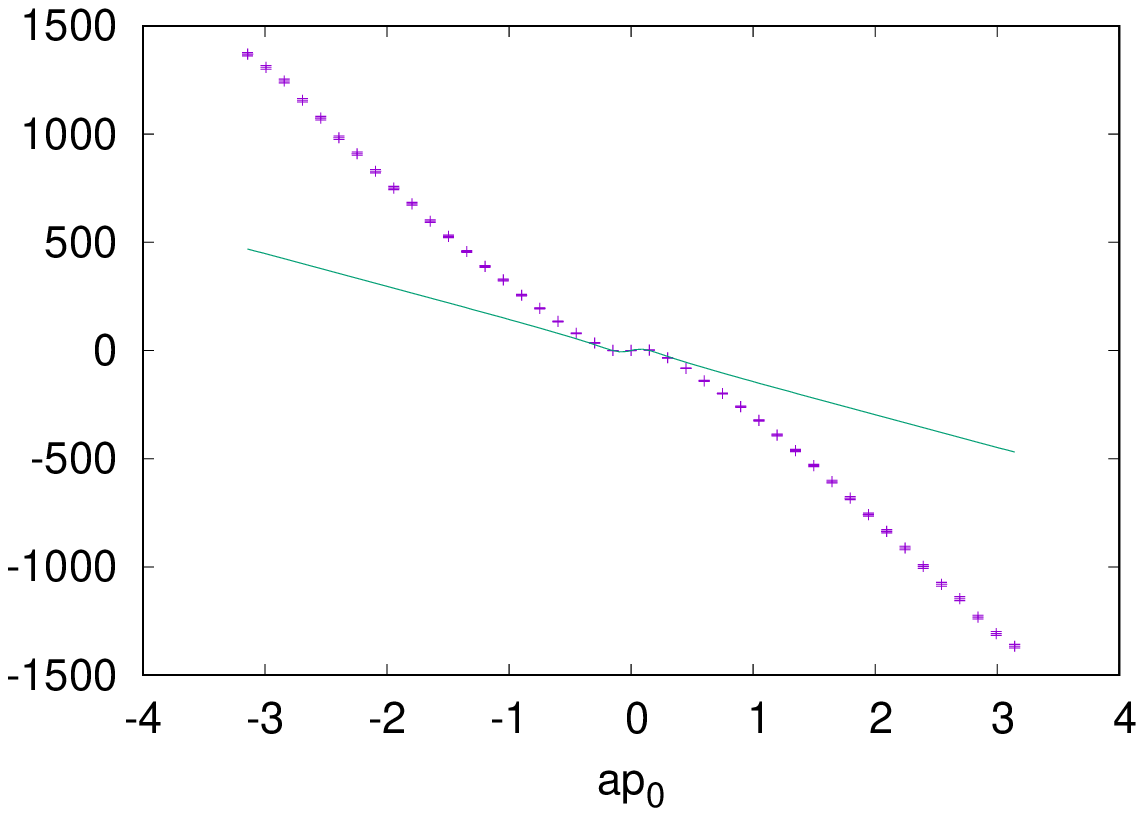}
    \caption{Imaginary part}
   \end{center}
  \end{subfigure}
 \end{center}
 \caption{$\langle T(p)T(-p)\rangle$ for~$D_4$, $L=42$, and~$p_1=\pi/21$.
 The fitting curve~\eqref{eq:emt.2} is depicted at once.}
 \label{fig:emt_spt-d_k3L42}
\end{figure}

\begin{figure}[ht]
 \begin{center}
  \begin{subfigure}{0.48\columnwidth}
   \begin{center}
    \includegraphics[width=\columnwidth]{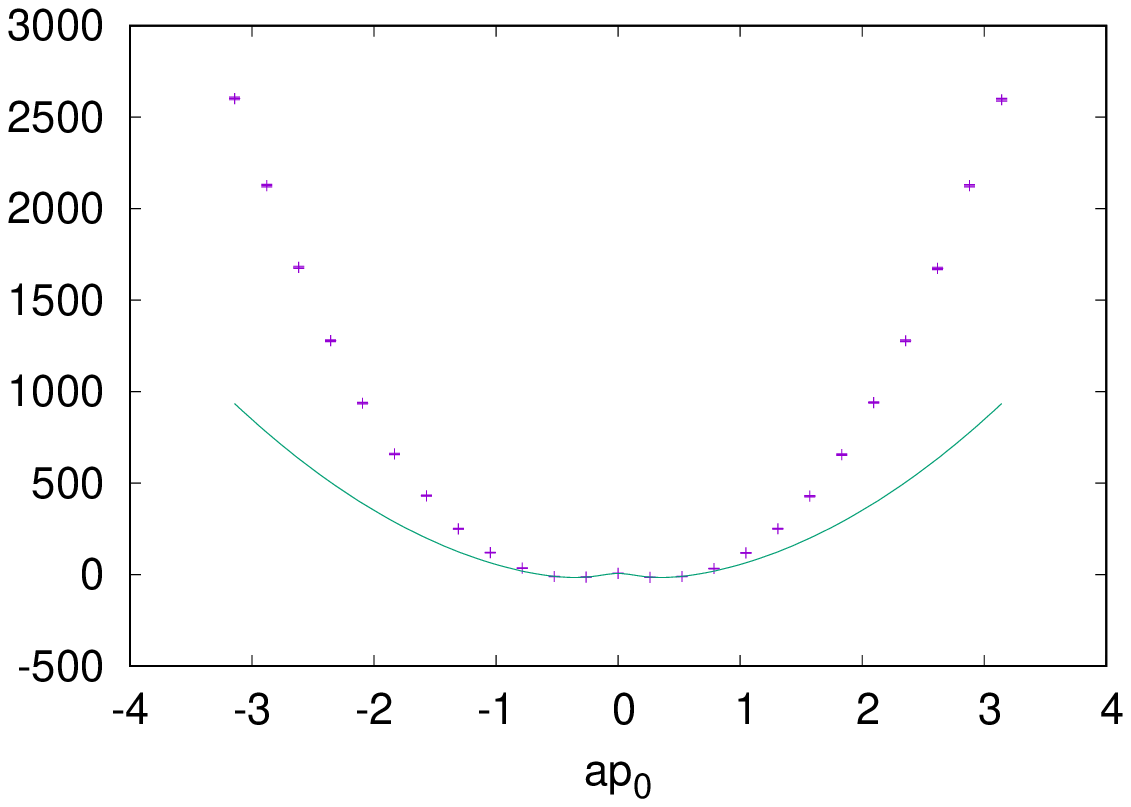}
    \caption{Real part}
   \end{center}
  \end{subfigure} \hspace*{1em}
  \begin{subfigure}{0.48\columnwidth}
   \begin{center}
    \includegraphics[width=\columnwidth]{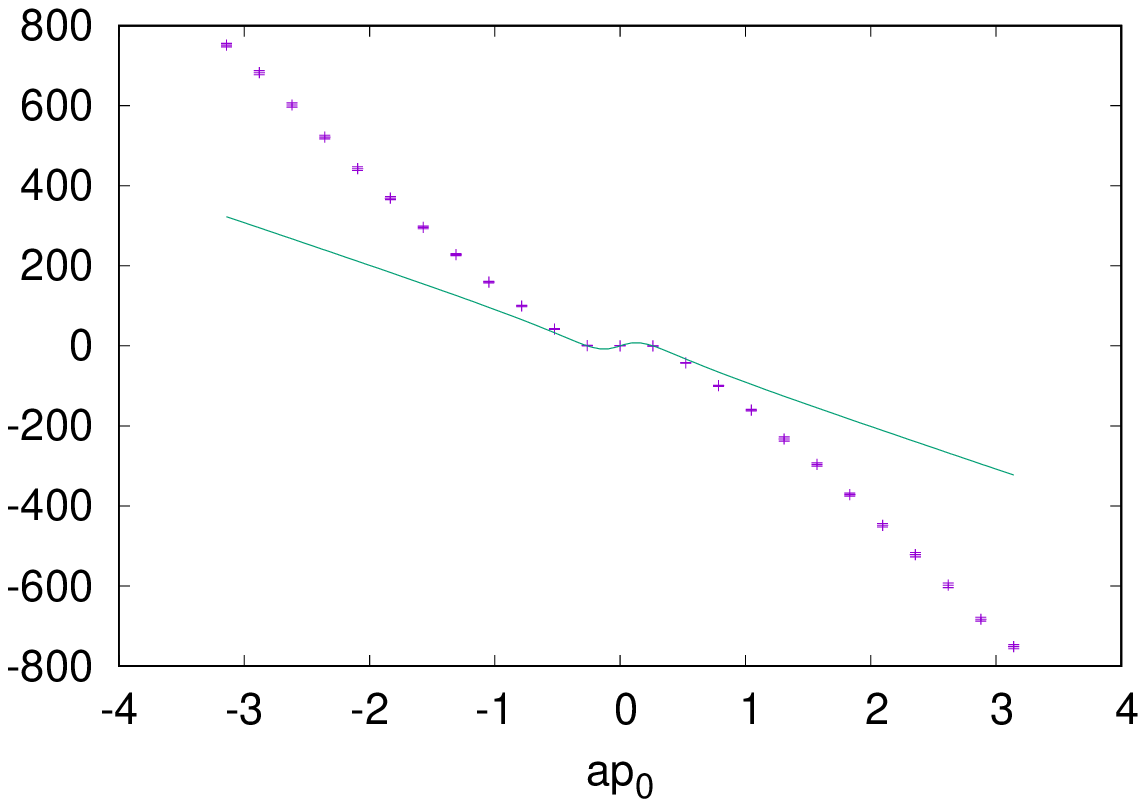}
    \caption{Imaginary part}
   \end{center}
  \end{subfigure}
 \end{center}
 \caption{$\langle T(p)T(-p)\rangle$ for~$E_7$, $L=24$, and~$p_1=\pi/12$.
 The fitting curve~\eqref{eq:emt.2} is depicted at once.}
 \label{fig:emt_spt-e_k3L24}
\end{figure}

We plot the correlation function~$\langle T(p)T(-p) \rangle$~\eqref{eq:emt.4}
in~Figs.~\ref{fig:emt_spt-d_k2L44}--\ref{fig:emt_spt-e_k3L24} for the maximal box size
with the fitting curve~\eqref{eq:emt.2};
the central charge~$c$ is obtained from the fit
in the IR region~$\frac{2\pi}{L} \leq |p| < \frac{4\pi}{L}$.
The left panel in each figure is devoted to the real part of the two-point function
and the right one is the imaginary part.
The horizontal axis indicates the momentum~$p_0$,
and the momentum~$p_1$ is fixed to the positive minimal value~$p_1 = 2\pi/L$.
In~Table~\ref{tab:res}, we tabulate the numerical results of the central charge
for all box sizes in the~$D_3$, $D_4$, and~$E_7$ models.

\begin{figure}[ht]
 \begin{center}
  \includegraphics[width=0.8\columnwidth]{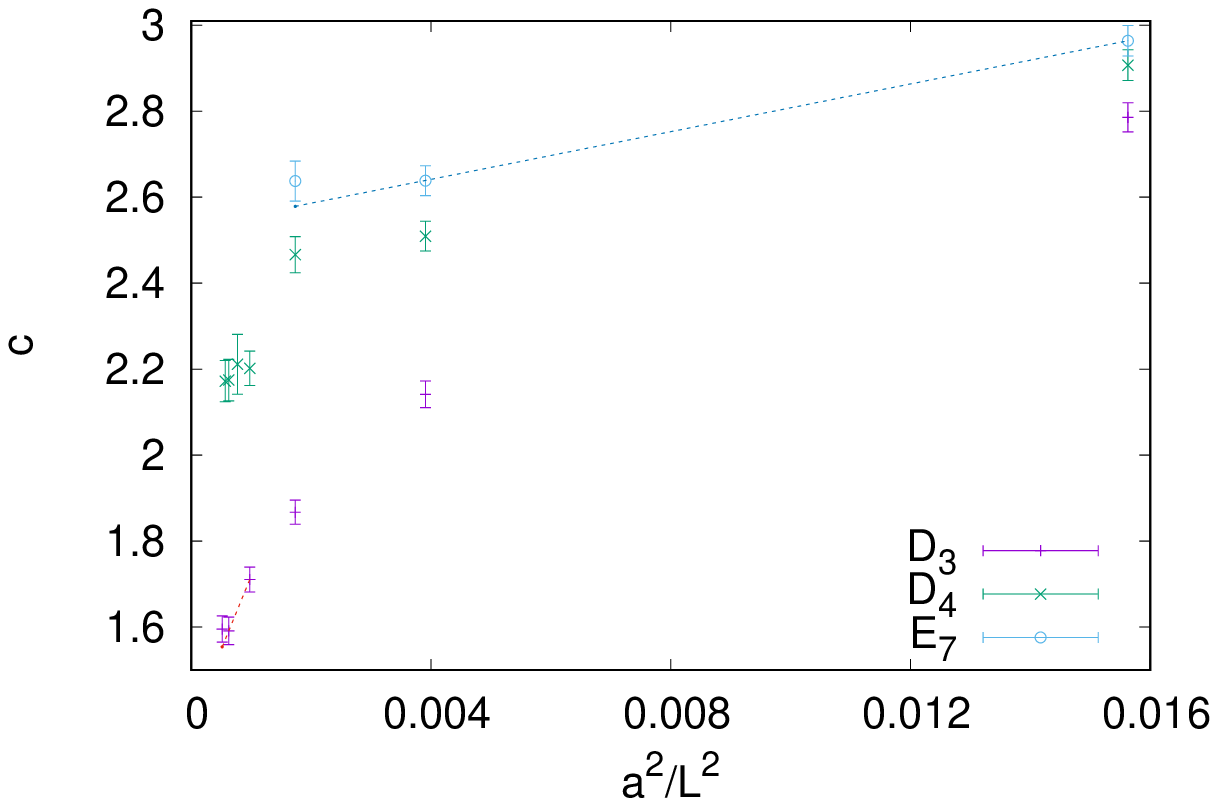}
 \end{center}
 \caption{Systematic error estimation for the central charge.}
 \label{fig:emt_sys}
\end{figure}

The central charge for the maximal box size in Table~\ref{tab:res} reads
\begin{align}
 c &= 1.595(31)(41) \qquad \text{for}\, D_3,
 \label{eq:c.1}\\
 c &= 2.172(48)(39) \qquad \text{for}\, D_4,
 \label{eq:c.2}\\
 c &= 2.638(47)(59) \qquad \text{for}\, E_7.
 \label{eq:c.3}
\end{align}
This is the main result in this paper.
Here, a number in the second parentheses indicates the systematic error
associated with the finite-volume effect.
In~Eqs.~\eqref{eq:c.1} and~\eqref{eq:c.3}, we estimate this as follows:
We pick out the largest three volumes for each minimal model;
from the central values at two smaller ones, we extrapolate to the larger~$L$ regime
as a linear function of the inverse volume~$1/L^2$, and then,
obtain an extrapolated value at the maximal volume (see~Fig.~\ref{fig:emt_sys});
the systematic error is identified with the deviation
between this and the central value in~Eqs.~\eqref{eq:c.1} and~\eqref{eq:c.3}.\footnote{%
The fit function~$\propto1/L^2$ is a possible choice,
but there would be no theoretical evidence to support this choice.
Because the behavior of the~$L\to\infty$ limit appears not quite smooth as
in~Fig.~\ref{fig:emt_sys}, we do not attempt to extrapolate to the infinite volume limit.
}
In~Eq.~\eqref{eq:c.2},
since we have more than two would-be convergent results at large~$L$,
the systematic error is estimated by the maximum deviation
of the central values at the three largest volumes as in~Ref.~\cite{Morikawa:2018ops}.
Eqs.~\eqref{eq:c.1}--\eqref{eq:c.3} are consistent with the expected values,
$1.5$ for~$D_3$ within~$\sim1.3\sigma$, $2$ for~$D_4$ within~$2\sigma$,
and~$2.666\dots$ for~$E_7$ within the numerical errors; the standard deviations
are evaluated by the sum of the statistical and systematic errors.


\begin{table}[ht]
 \centering
 \caption{The central charge obtained from the fit of the EMT correlator.
 The fitted momentum range is~$\frac{2\pi}{L}\leq|p|<\frac{4\pi}{L}$.}
 \begin{tabular}{lllll}\toprule
  Algebra & $L$  & $\chi^2/\text{d.o.f.}$ & $c$ & Expected value\\\midrule
  $D_3$   & $8$  & 6.056 & 2.786(34) & 1.5 \\
          & $16$ & 5.496 & 2.141(31) & \\
          & $24$ & 3.122 & 1.867(28) & \\
          & $32$ & 2.682 & 1.711(29) & \\
          & $40$ & 0.476 & 1.591(32) & \\
          & $44$ & 3.598 & 1.595(31) & \\
  \midrule
  $D_4$   & $8$  & 3.216 & 2.907(36) & 2 \\
          & $16$ & 3.738 & 2.509(34) & \\
          & $24$ & 1.946 & 2.466(42) & \\
          & $32$ & 2.832 & 2.202(40) & \\
          & $36$ & 1.109 & 2.211(70) & \\
          & $40$ & 2.276 & 2.175(48) & \\
          & $42$ & 1.177 & 2.172(48) & \\
  \midrule
  $E_7$   & $8$  & 2.220 & 2.964(36) & 2.666\dots \\
          & $16$ & 1.800 & 2.639(35) & \\
          & $24$ & 1.364 & 2.638(47) & \\
  \bottomrule
 \end{tabular}
 \label{tab:res}
\end{table}%

\begin{figure}[ht]
 \begin{center}
  \begin{subfigure}{0.48\columnwidth}
   \begin{center}
    \includegraphics[width=\columnwidth]{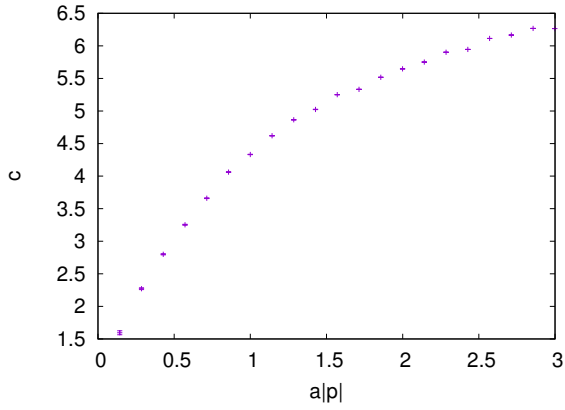}
    \caption{For~$D_3$ and~$L=44$.}
   \end{center}
  \end{subfigure} \hspace*{1em}
  \begin{subfigure}{0.48\columnwidth}
   \begin{center}
    \includegraphics[width=\columnwidth]{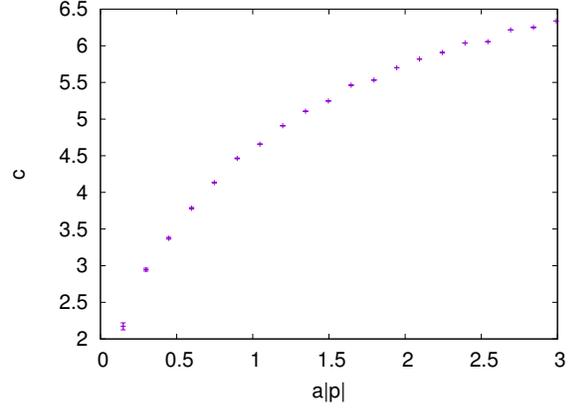}
    \caption{For~$D_4$ and~$L=42$.}
   \end{center}
  \end{subfigure}
  \begin{subfigure}{0.48\columnwidth}
   \begin{center}
    \includegraphics[width=\columnwidth]{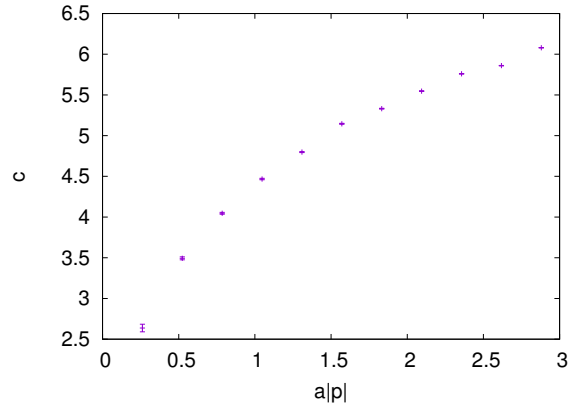}
    \caption{For~$E_7$ and~$L=24$.}
   \end{center}
  \end{subfigure}
 \end{center}
 \caption{``Effective central charge,''
 which changes as the function of~$|p|=\frac{2\pi}{L}n$ with fitted momentum regions,
 $\frac{2\pi}{L}n\leq|p|<\frac{2\pi}{L}(n+1)$, for~$n\in\mathbb{Z}_{+}$.}
 \label{fig:emt_fit}
\end{figure}

As mentioned in~Refs.~\cite{Kamata:2011fr,Morikawa:2018ops},
it is interesting to plot the ``effective central charge,'' which is analogous
to the Zamolodchikov's~$c$-function~\cite{Zamolodchikov:1986gt,Cappelli:1989yu}.
This is obtained from the fit in a variety of momentum regions from IR to UV;
we take the fitted momentum regions as
$\frac{2\pi}{L}n\leq|p|<\frac{2\pi}{L}(n+1)$ for $n\in\mathbb{Z}_{+}$.
Then Fig.~\ref{fig:emt_fit} shows that the ``effective central charge'' connects
the IR central charge to an UV value $c\approx6$.
This number is consistent with the central charge~$c=3N_{\Phi}$
in the expected free~$\mathcal{N}=2$ SCFT.
Recall that $N_{\Phi}$ is the number of supermultiplets
in the free~$\mathcal{N}=2$ WZ model, and we have set~$N_{\Phi}=2$ in this paper.

\clearpage
\section{Conclusion}
In this paper, we numerically studied the two-dimensional~$\mathcal{N}=2$
WZ model corresponding to the~$D_3$, $D_4$, and~$E_7$ minimal models.
Utilizing the supersymmetry-preserving formulation
with a momentum cutoff~\cite{Kadoh:2009sp},
we numerically determined the central charge from the IR behavior of the WZ model.
Although the theoretical background of our computational approach is not clear so far,
our results for the theories with two superfields are consistent
with the conjectured correspondence between the LG model and the minimal series of SCFT.
In the paper
and the preceding studies~\cite{Kawai:2010yj,Kamata:2011fr,Morikawa:2018ops},
we have the numerical evidences of typical minimal models:
the~$A_2$, $A_3$, $D_3$, $D_4$, $E_6$, and~$E_7$ models in Table~\ref{tab:1};
the~$A_4$ or~$E_8$~($\cong A_2\otimes A_4$) minimal model is left to be simulated.

To investigate superstring theory by using our numerical approach,
we may start from the numerical simulation of the~$A_4$ minimal model,
or simpler theories with several supermultiplets which is not a Gepner model.

\section*{Acknowledgments}
We would like to thank Katsumasa Nakayama, Hiroshi Suzuki,
and Hisao Suzuki for helpful discussions.
We are grateful to Hiromasa Takaura for a careful reading of the manuscript.
Our numerical computations were partially carried out
by using the supercomputer system ITO
of Research Institute for Information Technology (RIIT) at Kyushu University.
This work was supported by JSPS KAKENHI Grant Number JP18J20935.

\end{document}